\documentclass{article}
\usepackage{amssymb}

\usepackage{amsmath}

\input{tcilatex}

\begin{document}

\begin{tabular}{l}
\end{tabular}%
\vskip 3cm%

\begin{center}
{\Large Hamiltonian structure of thermodynamics with gauge}

\vskip 1cm%

{\large Roger Balian}

CEA/Saclay

Service de Physique Th\'{e}orique,

Orme des Merisiers, B\^{a}t. 774

F-91191 Gif sur Yvette Cedex

\vskip 0.3cm%

and

\vskip 0.3cm%

{\large Patrick Valentin}

Centre de recherches ELF

Chemin du Canal, BP 22, F-69360 Solaize

\vskip 2cm%
\end{center}

\underline{Abstract}

Denoting by $q^{i}\left( i=1,...,n\right) $ the set of extensive variables
which characterize the state of a thermodynamic system, we write the
associated intensive variables $\gamma _{i},$ the partial derivatives of the
entropy $S=S\left( q^{1},...,q^{n}\right) \equiv q_{0},$ in the form $\gamma
_{i}=-p_{i}/p_{0}$ where $p_{0}$ behaves as a gauge factor. When regarded as
independent, the variables $q^{i},p_{i}\left( i=0,...,n\right) $ define a
space $\mathbb{T}$ having a canonical symplectic structure where they appear
as conjugate.\ A thermodynamic system is represented by a $n+1$-dimensional\
gauge-invariant Lagrangean submanifold $\mathbb{M}$ of $\mathbb{T}.$\ Any
thermodynamic process, even dissipative, taking place on $\mathbb{M}$ is
represented by a Hamiltonian trajectory in $\mathbb{T},$ governed by a
Hamiltonian function which is zero on $\mathbb{M}.$ A mapping between the
equations of state of different systems is likewise represented by a
canonical transformation in $\mathbb{T}.$ Moreover a Riemannian metric
arises naturally from statistical mechanics for any thermodynamic system,
with the differentials $dq^{i}$ as contravariant components of an
infinitesimal shift and the $dp_{i}$'s as covariant ones. Illustrative
examples are given.

\newpage

\section{Introduction and outline}

In Callen's formulation of thermodynamics \cite{callen1}, the state of a
system is characterized at each time by the values of $n$ independent
quantities, usually extensive and conservative, that we shall denote as $%
q^{i}\left( i=1,...,n\right) .$ For a single fluid, these variables are the
internal energy $q^{1}\equiv U,$ the volume $q^{2}\equiv V$ and the numbers $%
q^{k}\equiv N^{k}\left( k=3,...,n\right) $ of molecules of each type $k;$
for a pure fluid, we have $n=3.$ For a composite system in local
equilibrium, the index $i$ is a compound index which\ refers both to each
such physical quantity and to each subsystem; for continuous media, these
subsystems are volume elements with sizes larger than the mean free path.\
All the equilibrium properties of a given system are embedded in the
expression of its entropy $S=S\left( q^{1},...,q^{n}\right) ,$ an additive,
extensive and concave function of the state variables $q^{i}.$ In
particular, the $n$ intensive variables $\gamma _{i}$ are defined as the
partial derivatives 
\begin{equation}
\gamma _{i}=\frac{\partial }{\partial q^{i}}S\!\left( q^{1},...,q^{n}\right)
.  \tag{1.1}
\end{equation}%
For a single fluid, they are expressed in terms of the temperature $T,$ the
pressure $P$ and the chemical potentials per particle $\mu _{k}$ as 
\begin{equation}
\gamma _{1}=\frac{1}{T},\quad \gamma _{2}=\frac{P}{T},\quad \quad \gamma
_{k}=-\frac{\mu _{k}}{T}\quad \left( k=3,...,n\right) .  \tag{1.2}
\end{equation}%
The relations (1.1) between the conjugate, extensive and intensive,
variables constitute the full set of equations of state.

In non-equilibrium situations, the variables $\gamma _{i}$ are the local
intensive variables. The fluxes which govern the dynamics in sufficiently
slow regimes are expressed in terms of differences of variables $\gamma _{i}$
for neighbouring subsystems.\ The simplest example is heat transfer across a
barrier separating two uniform\ thermal baths. In this case it is sufficient
to introduce $n=2$ variables, the energies $q^{1}\equiv U^{1}$ and $%
q^{2}\equiv U^{2}$ of the two baths, together with the associated
temperatures $\gamma _{1}=1/T_{1}$ and $\gamma _{2}=1/T_{2}.$ The dynamics
is governed by the\ conservation laws 
\begin{equation}
-\frac{dU^{1}}{dt}=\Phi =\frac{dU^{2}}{dt},  \tag{1.3}
\end{equation}%
and by the expression of the heat flux $\Phi $ across the barrier in terms
of the response coefficient $L\left( \gamma _{1},\gamma _{2}\right) $ which
characterizes the heat transport through this barrier,%
\begin{equation}
\Phi =L\left( \gamma _{1},\gamma _{2}\right) \left( \gamma _{2}-\gamma
_{1}\right) .  \tag{1.4}
\end{equation}

This formulation of thermodynamics is the one which arises naturally from
statistical mechanics \cite{callen1,balian1}, where the entropy $S$ is
identified with the missing information, or the disorder at microscopic
scale.\ Equilibrium is then characterized by looking for the largest
disorder, subject to constraints on the quantities $q^{i}.$ The variables $%
\gamma _{i}$ are the Lagrangean multipliers involved in this search. For
non-equilibrium thermodynamics microscopic foundations are also provided by
the maximum entropy method which yields the projection method in the limit
of negligible retardation effects (see, for instance, the review articles %
\cite{balian2} and \cite{rau}).

Although each state is characterized by the $n$ variables $q^{i}$ only, it
is convenient for practical purposes and for applications to make use of
additional variables. One is thus lead to introduce the $2n$-dimensional
space $\mathbb{\bar{T}}$ with coordinates $q^{1},...,q^{n},\gamma
_{1},...,\gamma _{n}.$ A thermodynamic system is represented in this space
by a $n$-dimensional manifold $\mathbb{\bar{M}}$ characterized by the set of 
$n$ equations of state (1.1); a state of this system is represented by a
point in $\mathbb{\bar{M}}.$ For a thermodynamic process, the dynamical
equations such as (1.1), (1.3), (1.4) couple the $2n$ time-dependent\
variables $q^{i}$ and $\gamma _{i}.$ It is here again natural to formulate
non-equilibrium thermodynamics in the space $\overset{-}{\mathbb{T}},$ where
a dynamical process is represented by a trajectory which is constrained by\
the equations of state (1.1) to lie in the manifold $\mathbb{\bar{M}}.$

Moreover, these equations of state are often written in an alternative form
where the entropy and the energy are interchanged.\ For a single fluid, the
function $S=S(U,V,N^{k})$ is thus inverted into $U=U(S,V,N^{k}).$ The
quantity $S$ can therefore be regarded as an additional extensive variable,
that we shall denote as $q^{0}$ so as to set all the variables $%
q^{0},q^{1},...,q^{n}$ on the same footing. For a given physical system
these $n+1$ variables are related to one another by%
\begin{equation}
q^{0}=S\!\left( q^{1},...,q^{n}\right) .  \tag{1.5}
\end{equation}%
The dissipation rate is the time-derivative $dq^{0}/dt.$ Altogether, the
thermodynamic variables $q^{0},q^{1},...,q^{n},\gamma _{1},...,\gamma _{n},$
now appear as the coordinates of a point in a $2n+1$-dimensional space $%
\overset{\sim }{\mathbb{T}}.$ In this space, which has a natural contact
structure [5--8], the dynamical equations involve the $n+1$ constraints
(1.1), (1.5). We have singled out the entropy rather than the energy\ among
the extensive variables because our results have a simpler theoretical
interpretation if we work in the entropy representation (1.5). For practical
purposes and for applications, it would be easy to transpose the formalism
into the energy representation, where $q^{0}$ and $q^{1}$ are interchanged
and where the associated intensive variables are $T,-P,\mu _{k}$ instead of
(1.2).

The consideration of the space $\overset{\sim }{\mathbb{T}}$ instead of the
initial $n$-dimensional configuration space $q^{1},...,q^{n}$\ has the
following interest in equilibrium thermodynamics. A physical system is
represented in this space $\overset{\sim }{\mathbb{T}}$ by a $n$-dimensional
manifold $\overset{\sim }{\mathbb{M}},$ characterized by the $n+1$ relations
(1.1), (1.5) or by any other equivalent set of $n+1$ equations of state.
(Strictly speaking $\overset{\sim }{\mathbb{M}}$ should be called a
submanifold since it is not defined intrinsically, but as a subset of $%
\overset{\sim }{\mathbb{T}}.)$ However experiments usually give only
indirect indications on the fundamental function $S\left(
q^{1},...,q^{n}\right) ,$ and are not sufficient to fully determine the
thermodynamic manifold $\overset{\sim }{\mathbb{M}}.$ Since less than the $%
n+1$ required equations of state are known, the remaining ones should be
determined by some phenomenological procedure.\ To this aim, it has been
advocated \cite{mrugala, benayoun1} to proceed by comparing the physical
system under study with some known system having the same degrees of freedom
but characterized by a different entropy function, or equivalently, by a
different thermodynamic potential.\ In the $2n+1$-dimensional thermodynamic
space $\overset{\sim }{\mathbb{T}}$ this system of reference is described by
a known manifold $\overset{\sim }{\mathbb{M}}_{0}.$ The incompletely known
manifold $\overset{\sim }{\mathbb{M}}$ to be determined is then deduced from 
$\overset{\sim }{\mathbb{M}}_{0}$ by setting up a correspondence between the
two systems, represented by some mapping $\varphi $ in $\overset{\sim }{%
\mathbb{T}}.$ The missing equations of state for $\overset{\sim }{\mathbb{M}}
$ follow from the corresponding ones for $\overset{\sim }{\mathbb{M}}_{0}.$
This mapping can be constructed by means of a continuous deformation scheme.
It is then generated through infinitesimal transformations in $\overset{\sim 
}{\mathbb{T}}.$ We shall regard the resulting motion as mock dynamics, and
shall interpret the deformation parameter as a fictitious time $\tau .$ The
method can be used to extrapolate, for instance, thermal properties from the
knowledge of the $P,V,T$ equation of state. In such problems we are
interested in the structure of a an imaginary dynamical flow which would
lead from $\overset{\sim }{\mathbb{M}}_{0}$ to $\overset{\sim }{\mathbb{M}},$
while the physical flow of non-equilibrium thermodynamics considered above
maps $\overset{\sim }{\mathbb{M}}$ onto itself.

The theoretical study of these mappings has already been achieved [8--10].
It is based on the remark that the physical manifolds $\overset{\sim }{%
\mathbb{M}}$ are not arbitrary.\ Although the relation (1.5) changes from
one manifold to another, there exists for each one some generating function $%
S$ which relates the conjugate variables $q^{i}$ and $\gamma _{i}\left(
i=1,...,n\right) $ according to (1.1). This is expressed in the $2n+1$%
-dimensional space $\overset{\sim }{\mathbb{T}}$ by introducing the 1-form 
\begin{equation}
\tilde{\omega}\equiv dq^{0}-\sum\limits_{i=1}^{n}\gamma _{i}\,dq^{i}, 
\tag{1.6}
\end{equation}%
which is non-degenerate at any point of $\overset{\sim }{\mathbb{T}}.$ Thus $%
\overset{\sim }{\mathbb{T}}$\ is endowed with a contact structure $(\overset{%
\sim }{\mathbb{T}},\tilde{\omega}\mathbb{)}$ [5--8].\ The existence of some
function $S\left( q^{1},...,q^{n}\right) $ such that all the equations
(1.1), (1.5) are satisfied is then equivalent to the vanishing of (1.6).
Among all the $n$-dimensional manifolds embedded in $\overset{\sim }{\mathbb{%
T}},$ the physical ones $\overset{\sim }{\mathbb{M}}$ must fulfil the
condition 
\begin{equation}
\tilde{\omega}=0  \tag{1.7}
\end{equation}%
for any infinitesimal variation of a state over $\overset{\sim }{\mathbb{M}},
$ an exterior differential equation which defines Legendre submanifolds.\ We
shall call them \textit{thermodynamic manifolds }to recall that the
vanishing of $\tilde{\omega}$ ensures the fulfilment of the thermodynamic
identities (1.1), (1.5). Further conditions on $\overset{\sim }{\mathbb{M}}$
are imposed by the extensivity and the concavity of entropy.

The transformations $\overset{\sim }{\mathbb{M}^{\prime }}\mathbb{=\varphi }%
\left( \overset{\sim }{\mathbb{M}}\right) $ of interest, which map in $%
\overset{\sim }{\mathbb{T}}$ a physical manifold either onto itself or onto
another, should preserve the condition (1.7) on the considered manifolds.
More generally, the transformations $\varphi $ are required to map \textit{%
any} thermodynamic manifold onto another thermodynamic manifold, that is, to
preserve the condition (1.7) \textit{everywhere.} This implies that they are
contact transformations in $\overset{\sim }{\mathbb{T}},$ which multiply the
1-form (1.6) by some non-vanishing function $\lambda $ of the coordinates $%
q^{0},q^{i},\gamma _{i}.$ More precisely the push-forward mapping $\varphi
_{\ast }=\varphi ^{-1\ast }$ which acts on the forms should satisfy 
\begin{equation}
\varphi _{\ast }\left( \tilde{\omega}\right) =\lambda \tilde{\omega}. 
\tag{1.8}
\end{equation}%
Applications of such contact transformations to thermodynamics have been
wor\-ked out [8--11]. In particular, by means of a continuous set of contact
transformations depending on a single deformation parameter $\tau $
interpreted as a fictitious time, one can explore the equilibrium
thermodynamic properties of a set of physically different systems labelled
by $\tau .$

The structure of the resulting equations of motion is somewhat similar with
that of Hamiltonian dynamics, though more complicated. However, it differs
from a symplectic\ Poisson structure, so that the extension of the standard
techniques of canonical Hamiltonian dynamics to the present problem is not
straightforward. The power of such techniques, which for instance readily
provide variational approaches, makes it desirable to modify the formulation
of thermodynamics so as to cast the equations of motion into a usual
Hamiltonian form. We would then benefit both from the flexibility brought in
by the idea of regarding the $2n+1$ variables $q^{0},q^{i},\gamma _{i}$ as
independent, and from the whole machinery of analytical mechanics.

We show below that this is feasible at little cost. We first introduce
(Section 2) a non-vanishing gauge variable $p_{0},$ without physical
relevance, which multiplies all the intensive variables. A new set of
variables $p_{i}$ is thus defined as $p_{i}=-p_{0}\gamma _{i}$ for $%
i=1,...,n.$ The $2n+1$-dimensional space $\overset{\sim }{\mathbb{T}}$ is
thereby extended into a $2n+2$-dimensional thermodynamic space $\mathbb{T}$
spanned by the variables $q^{i},p_{i}$ with $i=0,1,...,n.$ We associate
with\ a physical system a $n+1$-dimensional manifold $\mathbb{M}$ in $%
\mathbb{T}, $ parametrized for instance\ by the coordinates $q^{1},...,q^{n}$
and $p_{0}. $ A gauge transformation which changes the extra variable $p_{0}$
while keeping the ratios $p_{i}/p_{0}=-\gamma _{i}$ invariant is not
observable, so that a state of the system is represented by any point of a
one-dimensional ray lying in $\mathbb{M},$ along which the physical
variables $q^{0},q^{1},...,q^{n},\gamma _{1},...,\gamma _{n}$ are fixed.

We wish to study the transformations in the extended thermodynamic space $%
\mathbb{T}$ which map the thermodynamic manifolds $\mathbb{M}$ either onto
themselves (in non-equilib\-rium thermodynamics) or onto one another (in the
problem of generating equations of state). We show (Section 3) that, within
a suitable but irrelevant choice of gauge, these transformations are nothing
but canonical transformations of mechanics having some specific features.
While the variables $q^{i},\gamma _{i}\left( i=1,...,n\right) $ are
conjugate with respect to the entropy in the Legendre sense (1.1) of
thermodynamics, the variables $q^{i},p_{i}\left( i=0,1,...,n\right) $
moreover appear as canonically conjugate in the sense of Hamiltonian
dynamics. A symplectic structure is thus induced in the space $\mathbb{T}.$
An infinitesimal mapping among the manifolds $\mathbb{M}$ that preserves the
thermodynamic identities is represented by a (possibility time-dependent)
Hamiltonian, which generates a flow in $\mathbb{T}$ in terms of either the
real or the fictitious time, depending on the problem. The contact
transformations in $\overset{\sim }{\mathbb{T}}$ which relate the physical
thermodynamic manifolds $\overset{\sim }{\mathbb{M}}$ to one another are
recovered by elimination of the gauge variable $p_{0}.$

This relation between contact and canonical transformations is found here as
a direct outcome of the gauge invariance that we introduced in
thermodynamics.\ Actually mathematicians have recognized long ago \cite%
{herglotz1} that a contact structure of the type (1.6) in $2n+1$ dimensions
can be embedded into a symplectic structure in $2n+2$ dimensions by means of
the adjunction of an extra variable.\ This procedure, clearly presented by
Caratheodory \cite{caratheodory}, and later on termed as symplectization %
\cite{arnold2}, has a geometric interpretation in the theory of fibre
bundles.

On the other hand, exterior calculus, a standard technique in thermodynamics %
\cite{balian1}, has suggested to introduce in the $2n$-dimensional space of
extensive and intensive variables a symplectic structure which establishes a
duality between these variables. For instance, in the energy representation %
\cite{kijowski1}, one currently considers for a fluid the symplectic $2$%
-form $dT\wedge dS-dP\wedge dV$ in the corresponding $4$-dimensional space.
Likewise, in the entropy representation \cite{omohundro1}, the fundamental
symplectic $2$-form is $d\left( 1/T\right) \wedge dU+d\left( P/T\right)
\wedge dV.$ More generally, the $2n$-dimensional thermodynamics space $%
\mathbb{T}$ can be endowed with a symplectic structure generated by the $2$%
-form\ $\sum\limits_{i=1}^{n}d\gamma _{i}\wedge dq^{i}.$ The vanishing of
this form is known to characterize the surface $\overset{-}{\mathbb{M}}$ of $%
\overset{-}{\mathbb{T}}$ which describes the equations of state (1.1) of any
given thermodynamic system, so that $\overset{-}{\mathbb{M}}$ is a
Lagrangean submanifold of $\overset{-}{\mathbb{T}}.$ Deformations which map
thermodynamic systems onto one another in the $2n$-dimensional space $%
\overset{-}{\mathbb{T}}$ have also be considered \cite{janeczko}. They will
be recovered below (see Section 4) from canonical transformations in our $%
2n+2$-dimensional space $\mathbb{T}$ endowed with the symplectic form $%
\sum\limits_{i=0}^{n}dp_{i}\wedge dq^{i},$ as special cases when the entropy
variable $q^{0}$ may be left aside and when the gauge may be fixed as $%
p_{0}=-1.$ Our Hamiltonian dynamics in $\mathbb{T}$ thus generalizes and
encompasses both existing formulations in the $2n$- and $2n+1$-dimensional
spaces $\overset{-}{\mathbb{T}}$ and $\overset{\sim }{\mathbb{T}}.$

The present formalism applies in particular to the dynamical\ equations in
non-equilibrium thermodynamics (Section 4), which can thus be cast into a
Hamiltonian form.\ Given a set of equations of motion, it is usually not
obvious to recognize whether they have a Hamiltonian nature. This was done
for the hydrodynamics of ideal fluids \cite{morrison1}. Here we find another
type of Hamiltonian structure, for any dissipative system in the
thermodynamic regime. The existence of this structure is based on the idea
that any dynamical system can be embedded into a Hamiltonian system with a
double dimension (see ref.\ \cite{omohundro1}, Chap.\ 10). The remarkable
fact in non-equilibrium thermodynamics is the existence of a direct physical
interpretation for the additional variables.\ In case the entropy is left
aside, the evolution of the $n$ independent variables $q^{1},...,q^{n}$
which characterize the state of the system at each time has the same
structure as in analytical mechanics in spite of the existence of
dissipation.\ If we regard the set $q^{1},...,q^{n}$ as position variables,
their associated intensive variables $\gamma _{1},...,\gamma _{n}$ appear as
their conjugate momenta, and the dynamics is governed by an effective
Hamiltonian $f$ in the $2n$-dimensional space $\overset{-}{\mathbb{T}}.$ In
case the entropy in included among the dynamical variables, a similar
canonical structure is found provided we pay the price of introducing the
gauge variable $p_{0}$ upon which the effective Hamiltonian $h$ should
depend. Whereas the conjugate momentum of each $q^{i}$ is now $p_{i}$ for $%
i=1,...,n,$ the conjugate momentum of the\ entropy $q^{0}$ is $p_{0},$ and
the dissipation $dS/dt$ is expressed as $\partial h/\partial p_{0}.$ The
constraints (1.1), (1.5) satisfied at the initial time are conserved by the
Hamiltonian flow, which moreover implies that $f=0$ or $h=0$\ along the
physical trajectories. The fact that only the part of the Hamiltonian flow
located in the manifold (1.1), (1.5) is relevant for physics solves the
paradox of the existence of a Hamiltonian for dissipative dynamics. As an
illustration we work out in Section 4 Hamiltonian formulations for the
equations of motion (1.3), (1.4) describing heat transport. We show moreover
that these results are readily extended to any dissipative process in the
thermodynamic regime, whether mechanical, thermal, electrical or chemical.

The equations of state for different physical systems are also generated
from one another by means of Hamiltonian transformations in the space $%
\mathbb{T}.$ We illustrate this by writing in Section 5 Hamiltonian mappings
between different van der Waals fluids.

Finally we recall in Section 6 how quantum statistical mechanics generates %
\cite{balian2} a natural metric structure in thermodynamics \cite{weinhold1,
ruppeiner}. The thermodynamic Lagrangean manifolds $\overset{-}{\mathbb{M}}%
,\;\mathbb{M}$ or Legendrean manifolds $\mathbb{\tilde{M}}$ thus also
acquire a structure of Riemannian manifolds. The metric is obtained as 
\begin{equation}
ds^{2}=-\sum\limits_{i=1}^{n}d\gamma _{i}\,dq^{i}=\frac{1}{p_{0}}%
\sum\limits_{i=0}^{n}dp_{i}\,dq^{i},  \tag{1.9}
\end{equation}%
and the thermodynamically conjugate variables now appear as contravariant
and covariant coordinates on the manifold. The concavity of the entropy
function associated with $\mathbb{M}$ is equivalent to the positivity of $%
ds^{2}$ over $\mathbb{M}.$ This property may be used to select the
physically admissible manifolds, which should not only satisfy the algebraic
condition (1.7) but should also have a positive $ds^{2}$ so as to describe
thermodynamically stable systems.

\section{Thermodynamics as a gauge theory}

We start with a remark, drawn from statistical mechanics, about the
definition of the intensive variables. Consider, for instance, a fluid in
grand canonical equilibrium. In terms of the intensive variables $\gamma
_{i} $ defined by (1.2), its density operator $\hat{D}$ (in quantum
statistical mechanics) is expressed by 
\begin{equation}
\hat{D}\varpropto \exp \left[ -\frac{1}{k_{\mathrm{B}}}\left( \gamma _{1}%
\hat{H}+\sum\limits_{k=3}^{n}\gamma _{k}\,\hat{N}^{k}\right) \right] , 
\tag{2.1}
\end{equation}%
where $\hat{H}$ and $\hat{N}^{k}$ are the Hamiltonian and particle number
operators. The occurrence of Boltzmann's constant $k_{\mathrm{B}}$ arises
from the choice of the kelvin as the unit of temperature and the joule per
kelvin as the unit of entropy. We note from (1.2) that all the $\gamma _{i}$%
's are inversely proportional to the temperature. Thus, multiplying both $k_{%
\mathrm{B}}$ and the $\gamma _{i}\,^{\prime }s$ by a constant does not
affect the physics, since it amounts to changing the unit of temperature, or
equivalently the irrelevant coefficient in the definition of the von Neumann
entropy in quantum statistical mechanics, which is identified at equilibrium
with the absolute entropy of thermodynamics within a multiplicative constant.

This suggests us, for an arbitrary thermodynamic system, to introduce an
additional variable $p_{0}$ and to replace the set $\gamma _{i}$ by the new,
intensive, scaled variables 
\begin{equation}
p_{i}=-p_{0}\gamma _{i}\qquad \left( i=1,...,n\right) .  \tag{2.2}
\end{equation}%
In the resulting $2n+2$-dimensional space $\mathbb{T},$ a dilation of the
variables $q^{i},$ keeping the variables $p_{i}$ unchanged, is a physical
operation which leaves the equations of state (1.1), (1.5) unchanged in case
the system is extensive. Symmetrically we introduce a mathematical
operation, the dilation of the intensive variables, keeping the extensive
ones unchanged, 
\begin{equation}
p_{i}\mapsto \lambda p_{i},\qquad q^{i}\mapsto q^{i}\qquad \left(
i=0,1,...,n\right) ,  \tag{2.3}
\end{equation}%
where $\lambda $ is a non-zero constant. This operation, which does not
affect the physical variables $q^{0},...,q^{n},\gamma _{_{1}},...,\gamma
_{n},$ can be regarded as a \textit{gauge transformation} of the first kind.
In the trivial gauge $p_{0}=-1,$ we have $p_{i}=\gamma _{i}.$ A change of
gauge which leads to $p_{0}=-1/k_{\mathrm{B}}$ suppresses Boltzmann's
constant from the density operator (2.1) while changing the $\gamma _{i}$'s
into $p_{i}$'s$.$ (The macroscopic entropy remains, however, unchanged as
its expression in terms of $\hat{D}$ still contains a factor $k_{\mathrm{B}%
}.)$

We regard below, more generally, the dummy factor $p_{0}$ in (2.2) not as a
constant but as an independent variable. We thus allow gauge transformations
(2.3) of the second kind, where $\lambda $ is some function of the $2n+2$
coordinates $q^{i},p_{i}.$ An illustration will be given by Eq. (2.10),
which is obtained from the trivial gauge $p_{0}=-1$ by taking $\lambda
=p_{0}/p_{1}.$

In this formulation of thermodynamics as a gauge theory, the variables $%
p_{0},p_{1},$

\noindent $...,p_{n}$ are not completely meaningful physically, whereas
their ratios $\gamma _{i}$ given by (2.2) remain invariant in a gauge
transformation. The situation looks like classical electromagnetism, where
the e.m. potential is modified in a gauge transformation and is thus
physically unobservable (as the variables $p_{i}$), whereas the e.m. field
is well-defined (as the physical intensive variables $\gamma _{i}$). We
shall work in the $2n+2$-dimensional extended thermodynamic space $\mathbb{T}
$ with coordinates $q^{i},p_{i},$ including the gauge factor $p_{0}.$ This
is mathematically convenient, and physical interpretation will require to
return to the $2n+1$ physical coordinates, namely the entropy $q^{0},$ the $%
n $ extensive variables $q^{i}$ and the $n$ intensive variables $\gamma _{i}$
of Callen.

In the extended thermodynamic space $\mathbb{T},$ a state for a system is
not represented by a single point, but by any point of a one-dimensional ray
characterized by fixed values of the coordinates $q^{0},q^{1}...,q^{n}$ and
of the ratios $\gamma _{i}=-p_{i}/p_{0}.$ The points along this line,
equivalent for physics, result from one another by a gauge transformation
(2.3). A given system is described by a $n+1$-dimensional submanifold $%
\mathbb{M},$ which may be parametrized by the $n$ extensive variables $%
q^{1},...,q^{n}$ and by the gauge fixing variable $p_{0}.$ The entropy
function $S(q^{1},...,q^{n})$ of the system generates the remaining $n+1$
variables as

\begin{equation}
q^{0}=S(q^{1},...,q^{n}),\qquad p_{i}=-p_{0}\frac{\partial S\left(
q^{1},...,q^{n}\right) }{\partial q^{i}}.  \tag{2.4}
\end{equation}
We recalled in the introduction that the existence of such a generating
function $S$ characterizes the thermodynamic submanifolds $\overset{\sim }{%
\mathbb{M}}$ in the $2n+1$-dimensional space $\overset{\sim }{\mathbb{T}}.$
It is then expressed by the vanishing (1.7) of the $1$-form (1.6).
Accordingly, in the extended $2n+2$-dimensional space $\mathbb{T},$ we
introduce the 1-form

\begin{equation}
\omega \equiv \sum_{i=0}^{n}p_{i}dq^{i},  \tag{2.5}
\end{equation}
and the $n+1$-dimensional \textit{thermodynamic manifolds }$\mathbb{M}$ are
characterized by the vanishing of this form:

\begin{equation}
\omega =0\qquad \mathrm{over\ }\mathbb{M}.  \tag{2.6}
\end{equation}

The $1$-form $\omega $ induces a symplectic structure 
\begin{equation}
d\omega =\sum_{i=0}^{n}dp_{i}\wedge dq^{i}  \tag{2.7}
\end{equation}%
on $\mathbb{T}$ that we denote as $\left( \mathbb{T},d\omega \right) .$ Any
thermodynamic manifold $\mathbb{M}$ belongs to the set of the so-called
Lagrangean manifolds in $\mathbb{T},$ which are the integral submanifolds of 
$d\omega $ with maximum dimension $\left( n+1\right) .$ Moreover $\mathbb{M}$
is gauge invariant, which is implied by (2.6) but not by $d\omega =0.$

When the set $q^{i}$ contains all possible extensive variables, the
extensivity of the entropy function $S(q^{1},...,q^{n})$ is expressed by the
Gibbs--Duhem relation 
\begin{equation}
S=\sum_{i=1}^{n}\,q^{i}\frac{\partial S}{\partial q^{i}}.  \tag{2.8}
\end{equation}
Translating it into the space $\mathbb{T}$ by means of Eq.(2.4), we find

\begin{equation}
\sum_{i=0}^{n}\,p_{i\,}\,q^{i}=0.  \tag{2.9}
\end{equation}
This condition defines the $2n+1$-dimensional extensivity sheet $\mathbb{E}$
in the space $\mathbb{T},$ and the thermodynamic manifolds $\mathbb{M}$
should lie in this case within this surface $\mathbb{E}$. However, if the
systems are restricted to contain a fixed quantity of matter or to have a
fixed volume, the corresponding manifolds $\mathbb{M}$ are free from the
above constraint.

Statistical mechanics exhibits the different nature of the entropy $q^{0}$
and of the other extensive variables $q^{1},...,q^{n}.$ However it is
traditional in thermodynamics to introduce a symmetry between \ all $q^{i}$%
's for $i=0,1,...,n.$ In particular the energy as function of the entropy,
rather than the converse, is often used as a thermodynamic potential. This
symmetry is conveniently reflected in the present formalism by the use of
the gauged intensive variables $p_{0},p_{1},...,p_{n},$ as shown in (2.5),
(2.6) and (2.9). For instance, for a fluid, with the choice of gauge $%
p_{0}=-1,$ the intensive variables $p_{1},...,p_{n}$ reduce to the intensive
variables $\gamma _{i}$ arising from the entropy representation, given by
Eq.\thinspace (1.2). However, with the choice of gauge $p_{1}=-1,$ the other
intensive variables $p_{i},$ namely 
\begin{equation}
p_{0}=\frac{1}{\gamma _{1}}=T,\;\;p_{2}=-p_{0}\gamma
_{2}=-P,\;\;p_{k}=-p_{0}\gamma _{k}=\mu _{k}\;\;\left( k=3,...,n\right) , 
\tag{2.10}
\end{equation}
are those which arise from the energy representation of thermodynamics,
where the relation $q^{1}=U\left( q^{0},q^{2},...,q^{n}\right) $ replaces
(1.5).\thinspace Thus, the gauge transformation (2.3) with $\lambda
=p_{0}/p_{1}=-1/\gamma _{1}$ amounts to a switch from the entropy to the
energy representation.

\section{Canonical transformations in the extended \allowbreak configuration
space}

We proceed to study the transformations which, in the extended thermodynamic
space $\mathbb{T},$ map the thermodynamic manifolds $\mathbb{M}$ onto one
another. Such a mapping may describe a non-equilibrium thermodynamic
process, in which case the initial and the final manifolds are the same, or
it may describe a mapping between the equations of state of different
physical systems.

The thermodynamic submanifolds $\mathbb{M}$ satisfy the condition (2.6). We
impose that this condition should be \textit{preserved} by the considered
mappings $\mathbb{M}^{\prime }=\varphi (\mathbb{M)}$ \textit{for any }$%
\mathbb{M}.$ We recalled in Eq.\thinspace (1.8) that, in the $2n+1$%
-dimensional space $\overset{\sim }{\mathbb{T}}$ the corresponding mappings $%
\overset{\sim }{\mathbb{M}^{\prime }}=\varphi (\overset{\sim }{\mathbb{M}}%
\mathbb{)}$ should satisfy $\tilde{\omega}^{\prime }=\varphi _{\ast }(\tilde{%
\omega})=\lambda \omega $ where $\lambda $ is some non-vanishing function of
the coordinates. Likewise the mappings $\mathbb{M}^{\prime }=\varphi (%
\mathbb{M)}$ in the $2n+2$-dimensional space $\mathbb{T}$ that preserve the
thermodynamic identities should satisfy 
\begin{equation}
\omega ^{\prime }=\varphi _{\ast }(\omega )=\lambda \omega ,  \tag{3.1}
\end{equation}%
a property which ensures that $\omega =0$ implies $\omega ^{\prime }=0.$
However,\ the factor $\lambda $ in (3.1) can be absorbed by a gauge
transformation, since the multiplication of $\omega $ by a function $\lambda 
$ of the coordinates $q^{i},p_{i}$ can be achieved by the gauge
transformation (2.3) involving the same factor $\lambda .$ Hence, within an
irrelevant change of gauge in $\mathbb{M}^{\prime },$ any mapping $\mathbb{M}%
^{\prime }=\varphi (\mathbb{M)}$ which satisfies (3.1)\ is equivalent to a
transformation in $\mathbb{T}$ which conserves the $1$-form $\omega .$ We
can therefore take advantage of the gauge invariance and restrict ourselves
to the mappings $\mathbb{M}^{\prime }=\varphi (\mathbb{M)}$ such that 
\begin{equation}
\omega ^{\prime }=\varphi _{\ast }(\omega )=\omega .  \tag{3.2}
\end{equation}

The condition (3.2) implies that the $2$-form (2.7) obtained by taking the
exterior derivative of (2.5) is also conserved by the mapping $\varphi ,$
that is, 
\begin{equation}
\varphi _{\ast }(d\omega )=d\omega .  \tag{3.3}
\end{equation}%
This means that, if we endow the space $\mathbb{T}$ with a symplectic
structure with Poisson brackets 
\begin{equation}
\left\{ q^{i},p_{j}\right\} =\delta _{j}^{i},  \tag{3.4}
\end{equation}%
the considered mappings are canonical transformations of the same type as in
analytical mechanics \cite{caratheodory, arnold2, landau}.

Not every canonical transformation is admissible, however, since the
conservation (3.2) of the 1-form $\omega $ is a stronger condition that the
conservation (3.3) of the $2$-form $d\omega .$ Consider an infinitesimal
canonical transformation, generated by the (possibly time-dependent)
Hamiltonian $h(q^{0},q^{1},...,q^{n},p_{0},p_{1},...,p_{n}):$%
\begin{equation}
\dot{q}^{i}=\frac{\partial h}{\partial p_{i}},\qquad \dot{p}_{i}=-\frac{%
\partial h}{\partial q^{i}}.  \tag{3.5}
\end{equation}%
The evolution takes place either as function of the physical time $t$ in
non-equili\-brium thermodynamics, or as function of a fictitious time $\tau $
for continuous deformations relating different systems to one another.\ The
conservation of $\omega $ reads 
\begin{equation}
0=\sum\limits_{i=0}^{n}\left( p_{i}\,\,d\dot{q}^{i}+\dot{p}%
_{i}\,dq^{i}\right) =\sum_{i=0}^{n}\left( p_{i}\,d\frac{\partial h}{\partial
p_{i}}-\frac{\partial h}{\partial q^{i}}dq^{i}\right) =d\left(
\sum\limits_{i=0}^{n}p_{i}\frac{\partial h}{\partial p_{i}}-h\right) , 
\tag{3.6}
\end{equation}%
for any variation of the coordinates $q^{i},p_{i}\;(i=0,1,...,n).\,$As $h$
is defined within an additive constant, we can impose that it satisfies the
identity 
\begin{equation}
h\equiv \sum\limits_{i=0}^{n}p_{i}\frac{\partial h}{\partial p_{i}}, 
\tag{3.7}
\end{equation}%
or equivalently 
\begin{equation}
h\left( q^{0},q^{1},...,q^{n},\lambda p_{0},\lambda p_{1},...,\lambda
p_{n}\right) =\lambda h\left(
q^{0},q^{1},...,q^{n},p_{0},p_{1},...,p_{n}\right) ,  \tag{3.8}
\end{equation}%
where $\lambda $ is a constant or more generally a non-vanishing function.
The condition (3.8) also ensures that, if two points are deduced from each
other by a gauge transformation (2.3), their images in the evolution remain
related by the same transformation.\ The rays which represent the states in
the space $\mathbb{T}$ are thus deduced from one another in the Hamiltonian
motion, which therefore preserves the gauge invariance of the manifolds $%
\mathbb{M}$ in the dilation (2.3). Actually we have noted that the
thermodynamic manifolds $\mathbb{M}$ (for which $\omega =0)$\ are $n+1$%
-dimensional \textit{Lagrangean submanifolds} of $\mathbb{T}$ (for which $%
d\omega =0)$\ which are moreover \textit{gauge invariant.}

Altogether the mappings in the extended thermodynamic space $\mathbb{T}$
which preserve these two properties that characterize the thermodynamic
structure of the manifolds $\mathbb{M}$ are \textit{canonical
transformations }of analytical mechanics, generated by a \textit{Hamiltonian}
$h$ which is a \textit{homogeneous function }with degree 1 in the variables $%
p_{i}.$

If the set $q^{0},q^{1},...,q^{n}$ includes all the extensive variables, the
mappings should leave the extensivity sheet $\mathbb{E}$ defined by (2.9)
invariant. Expressing the time-derivative of (2.9) by means of (3.5) and
using (3.7), we obtain in this case%
\begin{equation}
h\equiv \sum\limits_{i=0}^{n}q^{i}\dfrac{\partial h}{\partial q^{i}}. 
\tag{3.9}
\end{equation}%
Hence, $h$ should also be a homogeneous function with degree 1 of the
variables $q^{i}.$ Note that, in spite of their formal analogy, the
identities (3.7) and (3.9) have different status, since dilation of the $%
p_{i}$'s is unphysical while dilation of the $q^{i}$'s is associated with
extensivity, and since (3.7) should always be satisified.

A finite canonical transformation may be characterized by its \textit{%
generating function} $\mathcal{H}(q^{0},...,q^{n},p_{0}^{\prime
},...,p_{n}^{\prime }),$ which produces the mapping $q^{i},p_{i}\longmapsto
q^{\prime i},p_{i}^{\prime }$ as 
\begin{equation}
p_{i}=\frac{\partial \mathcal{H}}{\partial q^{i}},\qquad q^{\prime i}=\frac{%
\partial \mathcal{H}}{\partial p^{\prime i}}\qquad (i=0,1,...,n).  \tag{3.10}
\end{equation}
(If the Jacobian of this mapping vanishes, one should use other generating
functions, where some initial variables $q^{i}$ are replaced by $p_{i}$'s,
and conversely for the final variables; see ref. \cite{arnold2}; an example
will be given by Eq. (5.8) below.) The conservation of the $1$-form $\omega $
is expressed, through the same calculation as in (3.6), by 
\begin{equation}
\mathcal{H}\equiv \sum\limits_{i=0}^{n}p_{i}^{\prime }\frac{\partial 
\mathcal{H}}{\partial p_{i}^{\prime }},  \tag{3.11}
\end{equation}
so that $\mathcal{H}$ is a homogeneous function of the variables $%
p_{i}^{\prime }$ with degree 1. Infinitesimal transformations (3.5) are
recovered from 
\begin{equation}
\mathcal{H}\approx \sum\limits_{i=0}^{n}p_{i}^{\prime }\,\,q^{i}+\varepsilon
h(q^{0},...,q^{n},p_{0}^{\prime },...,p_{n}^{\prime })  \tag{3.12}
\end{equation}
for $\varepsilon $ small. Note that the condition (3.11) results in the
vanishing of a Legendre transform of $\mathcal{H}$ with respect to the whole
set of variables $p_{i}^{\prime }.$ Hence no generating function depending
on the variables $q^{i},q^{\prime i}$ can produce here the Hamiltonian
mapping.\ Extensivity implies, as in (3.9), that $\mathcal{H}$ is a
homogenous function of the variables $q^{i}$ with degree 1. In this case,
the identity (3.11) implies that $\mathcal{H}$ vanishes when the points $%
(q^{i},p_{i})$ and $(q^{\prime i},p_{i}^{\prime })$ which are in
correspondence through (3.10) lie on the extensivity sheet.

We recover contact transformations in the space $\overset{\sim }{\mathbb{T}}$
by the projection $\pi $\ which\ eliminates the gauge factor $p_{0}$ from
Hamilton's equations, a procedure inverse from symplectization \cite%
{caratheodory, arnold2}. Indeed, the time-dependence of any function $%
g\left( q^{0},...,q^{n},\gamma _{1},...,\gamma _{n}\right) ,$ where the $%
\gamma _{i}$'s are expressed in terms of the canonical variables by (2.2),
is given, according to Hamilton's equations (3.5), by the usual Poisson
bracket 
\begin{equation}
\dot{g}=\left\{ g,h\right\} =\sum\limits_{i=0}^{n}\frac{\partial g}{\partial
q^{i}}\frac{\partial h}{\partial p_{i}}-\frac{\partial h}{\partial q^{i}}%
\frac{\partial g}{\partial p_{i}},  \tag{3.13}
\end{equation}%
generated by (3.4) in the space $\mathbb{T}.$ We can calculate it by using
the fact that 
\begin{equation}
f(q^{0},...,q^{n},\gamma _{1},...,\gamma _{n})\equiv -\frac{1}{p_{0}}h\left(
q^{0},...,q^{n},p_{0},...,p_{n}\right)   \tag{3.14}
\end{equation}%
is a function of the variables $\gamma _{i}\equiv -p_{i}/p_{0}$ only, as a
consequence of the homogeneity condition (3.8). The result, 
\begin{equation}
\dot{g}=\frac{\partial g}{\partial q^{0}}\left(
-f+\sum\limits_{i=1}^{n}\gamma _{i}\frac{\partial f}{\partial \gamma _{i}}%
\right) +\sum\limits_{i=1}^{n}\frac{\partial g}{\partial q^{i}}\frac{%
\partial f}{\partial \gamma _{i}}-\sum\limits_{i=1}^{n}\frac{\partial g}{%
\partial \gamma _{i}}\left( \frac{\partial f}{\partial q^{i}}+\gamma _{i}%
\frac{\partial f}{\partial q^{0}}\right) ,  \tag{3.15}
\end{equation}%
generates the known contact flow \cite{mrugala} in the thermodynamic space $%
\overset{\sim }{\mathbb{T}}.$ The greater simplicity of the canonical flow
in the extended space $\mathbb{T}$ arises from the mock dynamics 
\begin{equation}
\dot{p}_{0}=-\frac{\partial h}{\partial q^{0}}  \tag{3.16}
\end{equation}%
that we have introduced for the gauge variable, which allowed us to get rid
of the factor $\lambda $ in (3.1), whereas contact transformations should
manage with this factor in (1.8). We can thus regard a contact
transformation in $\overset{\sim }{\mathbb{T}}$ as a Hamiltonian
transformation in $\mathbb{T}$ followed by the gauge transformation which
brings $p_{0}$ to its initial value $-1.$ However, the latter gauge
transformation (2.3) cannot in general be generated by a Poisson structure,
so that this structure is lost in the elimination of $p_{0}.$ As it often
occurs in gauge theories, the equations become more complicated when the
gauge is fixed without care.

Note that the antisymmetry of the Poisson bracket (3.13), which generates
the motion in the $2n+2$-dimensional space $\mathbb{T},$ is lost when this
motion is projected as (3.15) on the $2n+1$-dimensional space $\overset{\sim 
}{\mathbb{T}}.$ Indeed the first term $-\left( \partial g/\partial
q_{0}\right) f$ of (3.15) has no counterpart.\ We can trace back this lack
of symmetry to a hidden difference which already existed between $g$ and $h$
in the Poisson bracket (3.13). Being a physical quantity, $g$ is a
homogeneous function of the variables $p_{i}$ with degree $0.$ However, $h,$
the generator of the motion in the space $\mathbb{T},$ is not a function of
the physical variables $q^{i},\gamma _{i}$ only, but it includes $p_{0}$
and\ is homogeneous in the $p_{i}$'s with degree $1.$ The first term of
(3.15) arises from this difference of behaviour of $g$ and $h$ in a gauge
transformation.

We can represent the relations between the contact space $(\overset{\sim }{%
\mathbb{T}},\tilde{\omega})$ and the symplectic space $(\mathbb{T},d\omega )$
by the following diagram: 
\begin{equation*}
\begin{array}{ccc}
\qquad \qquad \qquad \left( \mathbb{T},d\omega \right)  & ^{\underrightarrow{%
\text{{\large symplectic\ transformation,}\textrm{\ \ }}h}} & \left( \mathbb{%
T},d\omega \right) \qquad \qquad \qquad  \\ 
\!\!\!\mathrm{Symplectization}\left\uparrow 
\begin{array}{c}
\end{array}%
\right.  &  & \left\downarrow 
\begin{array}{c}
\end{array}%
\right. \text{\textrm{Projection, }}\pi  \\ 
\qquad \qquad \qquad (\overset{\sim }{\mathbb{T}},\tilde{\omega}) & _{%
\overrightarrow{\text{ \ {\large contact\ transformation,}\ \textrm{\ }}f%
\text{ \ }}} & (\overset{\sim }{\mathbb{T}},\tilde{\omega})\qquad \qquad
\qquad 
\end{array}%
\end{equation*}%
The symplectization, usually performed \cite{arnold2} by constructing $%
\mathbb{T}$ as a fibre bundle associated with the configuration space $%
q^{0},...,q^{n},$ amounts here to the introduction of a gauge structure in
the space of intensive variables $\gamma _{i}=-p_{i}/p_{0},$ whereas the
projection $\pi $ amounts to the fixation of the gauge through $p_{0}=-1.$
The Hamiltonians $h$ which are equivalent to contact transformations $f$ are
constrained by the homogeneity condition (3.8), which ensures that the gauge
invariance of the Lagrangean manifolds $\mathbb{M}$ is preserved.

\section{Hamiltonian equations for dissipative dynamics and quasi-static
processes}

The above formalism holds in particular for a continuous mapping which
leaves some physical manifold $\mathbb{M}_{0}$ invariant. The equations of
motion (3.5) can thus describe non-equilibrium thermodynamic processes which
take place in a given system characterized by $\mathbb{M}_{0},$ and which
are represented by the motion of a point in $\mathbb{M}_{0}.$ In these
dynamics, the other thermodynamic manifolds $\mathbb{M}$ are not kept
invariant but are transformed into one another, contrary to $\mathbb{M}_{0}.$
The effective Hamiltonian $h$ will therefore depend on the equations of
state (2.4) which parametrize $\mathbb{M}_{0}$ (or equivalently on the
entropy function $S$ of the system), as well as on the transport
coefficients. This effective Hamiltonian $h$ should not be confused with the
microscopic Hamiltonian $\hat{H}$ which governs the dynamics in statistical
mechanics.

Apart from its general property (3.8), $h$ should here be such that the
Hamiltonian flow that it generates through (3.5) lies, for any point of $%
\mathbb{M}_{0},$ in $\mathbb{M}_{0}$ itself. In particular, the relation $%
q^{0}=S(q^{1},...,q^{n})$ should be conserved, which is expressed by writing
its time-derivative as 
\begin{equation}
\frac{\partial h}{\partial p_{0}}=\sum\limits_{i=1}^{n}\frac{\partial S}{%
\partial q^{i}}\frac{\partial h}{\partial p_{i}}=-\frac{1}{p_{0}}%
\sum\limits_{i=1}^{n}p_{i}\frac{\partial h}{\partial p_{i}}.  \tag{4.1}
\end{equation}%
Together with (3.7), this implies that 
\begin{equation}
h=0\qquad \mathrm{over}\;\mathbb{M}_{0}.  \tag{4.2}
\end{equation}%
Conversely, the general homogeneity property (3.8) of $h$ and the \textit{%
vanishing of }$h$ \textit{on the manifold }$\mathbb{M}_{0}$ which represents
the given system in the extended thermodynamic space $\mathbb{T}$ are
sufficient to ensure that the \textit{equations of state} (2.4) \textit{%
remain satisfied at all times }provided they are satisfied at the initial
time.\ This is readily checked by regarding the set (2.4) as a
parametrization of $\mathbb{M}$ in terms of the independent variables $%
q^{1},...,q^{n},p_{0}.$ This can also be seen as a consequence of (4.2) and
of the conservation, along the motion, of the $1$-form $\omega $ which thus
remains zero. The vanishing of $h$ over $\mathbb{M}_{0}$ is the counterpart
of the vanishing of $f$ in the corresponding contact transformation in $%
\overset{\sim }{\mathbb{T}}$ that keeps $\overset{\sim }{\mathbb{M}}_{0}$\
invariant, a property rigorously proven in ref. \cite{mrugala}.

One among the equations of motion, 
\begin{equation}
\dot{q}^{0}=\frac{\partial h}{\partial p_{0}},  \tag{4.3}
\end{equation}
describes the rate of change of the entropy, which for an isolated system is
the \textit{dissipation}. Thus, although the variable $p_{0}$ has in itself
no physical meaning, its occurrence in the effective Hamiltonian is
essential to deal with dissipative processes. It would clearly not have been
possible to assign to an irreversible process Hamiltonian equations
including the time-derivative $\dot{q}_{0}$ of entropy without introducing
such an extra variable. Dynamical features which apparently contradict usual
properties of Hamiltonian motions, such as the fact that the dissipation
(4.3) cannot be negative, or the convergence at large times of the
trajectories towards a fixed point which describes global equilibrium of the
system, arise from a special choice of the initial point, which should
always lie on $\mathbb{M}_{0}.$ In particular, the Liouville theorem is
compatible with the convergence of physical trajectories towards a fixed
point,\ because there exist neighbouring unphysical trajectories, close to $%
\mathbb{M}_{0}$ but outside it, which diverge away from this fixed point. We
shall illustrate this fact in Eq.\thinspace (4.13) below.

Since the Hamiltonian equations (3.5) are here physically meaningful only
for the flow in $\mathbb{M}_{0},$ the Hamiltonian $h$ is not defined in a
unique fashion for a given physical process. Changes of $h$ which do not
modify its value $(h=0)$ on $\mathbb{M}_{0}$ and its first-order derivatives
on $\mathbb{M}_{0}$ are irrelevant for physics.

As an illustrative example, let us consider heat transfer between two
thermal baths. We assume the thermal conductivity of each bath to be much
larger than that of the barrier, so that their two energies $U^{1}$ and $%
U^{2}$ are sufficient to characterize at each time the state of the system.\
The entropy $S$ is the sum of the entropies $S_{1}\left( U^{1}\right) $ and $%
S_{2}\left( U^{2}\right) $ of the two baths.\ We recalled in the
introduction the equations of motion (1.3), (1.4) for this model. In the $6$%
-dimensional extended thermodynamic space $\mathbb{T},$ the manifold $%
\mathbb{M}_{0},$ parametrized by $q^{1}=U^{1},q^{2}=U^{2}$ and the gauge
variable $p_{0},$ is characterized by the equations 
\begin{equation}
q^{0}=S_{1}(q^{1})+S_{2}(q^{2}),\quad p_{1}=-p_{0}\beta _{1}(q^{1}),\quad
p_{2}=-p_{0}\beta _{2}(q^{2}),  \tag{4.4}
\end{equation}
where we denote as 
\begin{equation}
\beta _{1}\left( U_{1}\right) =\frac{dS_{1}\left( U_{1}\right) }{dU_{1}}%
,\qquad \beta _{2}\left( U_{2}\right) =\frac{dS_{2}\left( U_{2}\right) }{%
dU_{2}}  \tag{4.5}
\end{equation}
the inverse temperatures of the baths regarded as functions of their
energies. We readily check that on the manifold (4.4) the equations of
motion (1.3),(1.4), rewritten in terms of $q^{i},p_{i}\;(i=0,1,...,n),$ can
be generated through (3.5) from the Hamiltonian 
\begin{equation}
h=L(\beta _{1},\beta _{2})\left[ -\frac{1}{2p_{0}}(p_{1}-p_{2})^{2}+\frac{%
p_{0}}{2}(\beta _{1}-\beta _{2})^{2}\right] ,  \tag{4.6}
\end{equation}
which depends on $q^{1}$ and $q^{2}$ through $\beta _{1}(q^{1})$ and $\beta
_{2}(q^{2}).$ As it should, $h$ vanishes on the physical manifold (4.4), and
is of degree 1 in the variables $p_{i}.$ The latter property ensures that
the dissipation (4.3) is the one, $\sum\limits_{i=1}^{n}\gamma _{i}\,\dot{q}%
^{i},$ associated with the flux (1.4). The conservation of energy is
reflected by the fact that $h$ depends on $p_{1}$ and $p_{2}$ only through
their difference.

As noted above, we can alternatively use for $h$ any Hamiltonian equivalent
to (4.6) to first-order in the variables $p_{1}+p_{0}\,\beta
_{1}(q^{1}),p_{2}+p_{0}\beta _{2}(q^{2})$ and $q^{0}-S(q^{1},q^{2})$ which
vanish on the manifold $\mathbb{M}_{0}$. For instance, we can take, in terms
of the flux $\Phi (\gamma _{1},\gamma _{2})=L(\gamma _{1},\gamma _{2})\left(
\gamma _{2}-\gamma _{1}\right) $ of energy escaping from the bath 1, which
is a function of the intensive variables, the Hamiltonian 
\begin{equation}
h=\Phi (-\frac{p_{1}}{p_{0}},-\frac{p_{2}}{p_{0}})\left[ (p_{1}+p_{0}\beta
_{1})-(p_{2}+p_{0}\beta _{2})\right] .  \tag{4.7}
\end{equation}

The Hamiltonians (4.6) or (4.7) do not depend on the entropy variable $q^{0},
$ so that the time-derivative 
\begin{equation}
\dot{p}_{0}=-\frac{\partial h}{\partial q^{0}}  \tag{4.8}
\end{equation}%
of the gauge variable $p_{0}$ vanishes. The elimination of the gauge is
trivial in such a situation. By fixing the gauge so that $p_{0}=-1$ at all
times, and by keeping aside the entropy variable $q^{0},$ we can then regard
the $2n$-dimensional space $\overset{-}{\mathbb{T}}$ spanned by $%
q^{1},...,q^{n},\gamma _{1},...,\gamma _{n}$ as a mechanical phase space in
which the extensive and intensive variables $q^{i}$ and $\gamma
_{i}\,(i=1,...,n)$ appear as canonically conjugate.\ Their dynamics are
generated by the Hamiltonian $p_{0}f=-f$ defined by (3.14), as 
\begin{equation}
\dot{q}^{i}=\frac{\partial f}{\partial \gamma _{i}},\qquad \dot{\gamma}_{i}=-%
\frac{\partial f}{\partial q^{i}}.  \tag{4.9}
\end{equation}%
Actually Eq. (3.14) defines in general a contact Hamiltonian $f$\ in the $%
2n+1$-dimensional space $\overset{\sim }{\mathbb{T}},$ but the absence of $%
q^{0}$ allows us to regard $f$ as an ordinary Hamiltonian in the $2n$%
-dimensional symplectic space $\overset{-}{\mathbb{T}},$ in the special case
when the effective Hamiltonian in the space $\mathbb{T}$ does not depend on
the entropy variable $q^{0}.$ For the heat transfer problem, the Hamiltonian 
$f$ which arises from (3.14) and (4.6) is, in terms of the ``position''
variables $q^{1},q^{2}$ and their conjugate ``momenta'' $\gamma _{1},\gamma
_{2},$%
\begin{equation}
f=\tfrac{1}{2}L\left( \beta _{1},\beta _{2}\right) \left[ \left( \gamma
_{1}-\gamma _{2}\right) ^{2}-\left( \beta _{1}-\beta _{2}\right) ^{2}\right]
.  \tag{4.10}
\end{equation}%
Equivalently as regards the physical dynamics, we get from (4.7) 
\begin{equation}
f=\Phi \left( \gamma _{1},\gamma _{2}\right) \left[ \left( \gamma _{1}-\beta
_{1}\right) -\left( \gamma _{2}-\beta _{2}\right) \right] .  \tag{4.11}
\end{equation}

The Hamiltonians (4.6) in $\mathbb{T}$ or (4.10) in $\overset{-}{\mathbb{T}}$
involve two terms, the first one of ``kinetic'' type, the second one of
``potential'' type. These two terms occur with opposite signs, in contrast
to what usually happens in mechanics. For the motions in which we are
interested, they exactly cancel out at the initial time, and hence at all
times. As indicated above, the special choice of an initial point in $%
\mathbb{M}_{0}$ prevents the divergence of trajectories which is expected
from the wrong sign of the potential term. We illustrate this point by fully
solving the equations (4.9) in the whole space $\mathbb{T}$ for a model
where $L$ is a constant and where the entropies are quadratic functions of
the energies: 
\begin{equation}
S_{i}\left( U^{i}\right) =a_{i}U^{i}-\tfrac{1}{2}b_{i}\left( U^{i}\right)
^{2},\qquad \left( i=1,2\right) .  \tag{4.12}
\end{equation}%
(The Gibbs--Duhem homogeneity condition (2.8) is not satisfied here because
only one extensive variable, $q^{i}=U^{i},$ has been introduced for each
subsystem; hence $b_{i}^{-1}$ is an extensive quantity.) The Hamiltonian
(4.10) is that of a two-dimensional harmonic oscillator in which the sign of
the potential $-\frac{1}{2}L\left( \beta _{1}-\beta _{2}\right) ^{2}=-\frac{1%
}{2}L\left( a_{1}-b_{1}q^{1}-a_{2}+b_{2}q^{2}\right) ^{2}$ has been
inverted. Its general flow is 
\begin{equation}
\gamma _{1}-\gamma _{2}=A\mathrm{e}^{\Gamma t}+B\mathrm{e}^{-\Gamma
t},\qquad \beta _{1}-\beta _{2}=-A\mathrm{e}^{\Gamma t}+B\mathrm{e}^{-\Gamma
t},  \tag{4.13}
\end{equation}%
where $\Gamma \equiv \left( b_{1}+b_{2}\right) L,$ with the total energy $%
q^{1}+q^{2}$ and the quantity $b_{2}\gamma _{1}+b_{1}\gamma _{2}$ as
constants of the motion. The concavity of the entropy functions (4.12)
implies $b_{i}>0,$ and the positivity of the response coefficient $L$ then
implies $\Gamma >0.$ The arbitrary constants $A\;$and $B$ are determined by
the initial conditions, which for a physical process entail $\gamma
_{1}-\gamma _{2}=\beta _{1}-\beta _{2}$ and hence $A=0.$ Thus only the
decaying terms $B\mathrm{e}^{-\Gamma t}$ occur in (4.13) for physical
processes, although the diverging terms $A\mathrm{e}^{\Gamma t}$ are present
for non-physical trajectories which lie outside the manifold $\mathbb{M}_{0}.
$ Likewise the dissipation evaluated from (4.12) and (4.13), $\dot{S}%
=L\left( B^{2}\mathrm{e}^{-2\Gamma t}-A^{2}\mathrm{e}^{2\Gamma t}\right) ,$
remains positive at all times only on $\mathbb{M}_{0}.$ It is therefore
important, in order to use Hamilton's equations (3.5) or (4.9) for a
numerical solution of dissipative motions, to enforce the constraints (2.4)
which characterize $\mathbb{M}_{0},$ although these constraints are
conserved along the motion, because small errors may produce increasing
spurious effects.

In spite of its simplicity, the heat transfer problem considered above is a
prototype for \textit{any process} of non-equilibrium thermodynamics,
describing transfer of heat, of momentum, of particles, or chemical
reactions. In any such case, the time-derivative of each extensive variable $%
q^{i},$ where the compound index $i=k,\alpha $ refers both to the nature $k$
of the quantity transferred and to the considered subsystem $\alpha ,$ is
expressed as a sum of outgoing fluxes $-\Phi _{\beta }^{i}\left( \gamma
_{1},...,\gamma _{n}\right) ,$ which depend on the intensive variables $%
\gamma _{i}$\ and which involve the subsystems $\beta $ that interact with $%
\alpha .$ The conservation laws are expressed by $\Phi _{\beta }^{k\alpha
}=-\Phi _{\alpha }^{k\beta }.$ If we keep aside the entropy variable $q^{0}$
and$\ $fix the gauge as $p_{0}=-1,$ we can generate the dynamics in the $2n$%
-dimensional space $\overset{-}{\mathbb{T}}$\ from the effective Hamiltonian 
\begin{equation}
f=\sum_{i,\beta }\Phi _{\beta }^{i}\left( \gamma _{1},...,\gamma _{n}\right) 
\left[ \frac{\partial S\left( q^{1},...,q^{n}\right) }{\partial q^{i}}%
-\gamma _{i}\right] .  \tag{4.14}
\end{equation}
This expression generalizes (4.11), which involved only two opposite fluxes $%
\Phi ^{1}=\Phi =-\Phi ^{2}.$ It also applies to non-isolated systems, in
which case some fluxes are imposed by external sources.\ Since the
trajectories of interest satisfy the constraints $\gamma _{i}=\partial
S/\partial q^{i},$ the Hamiltonian flow in the whole space can be stabilized
around those constraints by adding to (4.14) terms proportional to 
\begin{equation}
\left[ \frac{\partial S\left( q^{1},...,q^{n}\right) }{\partial q^{i}}%
-\gamma _{i}\right] ^{2},  \tag{4.15}
\end{equation}
which do not affect the dynamics on the physical manifold $\mathbb{M}.$

Effective Hamiltonians such as (4.14) are not arbitrary.\ They should
satisfy several properties imposed by the theory of non-equilibrium
thermodynamics \cite{callen1,balian1}. The entropy function $S\left(
q^{1},...,q^{n}\right) $ should be concave, expressing \textit{stability of
matter.} The fluxes should be two by two opposite, so as to ensure the 
\textit{conservation} laws.\ The matrix of response coefficients\textit{\ }$%
L $ should be positive so as to ensure that the \textit{dissipation }is not
negative. Finally the symmetry and invariance laws should be reflected in
the form of the functions $S$ and $\Phi .$

In the special case of \textit{quasi-static} processes, there is no
dissipation.\ Such processes are usually considered for a non-isolated
system which remains nearly in equilibrium at each time under the effect of
external sources. If it does not interact with a heat source, its entropy
remains constant, so that the effective Hamiltonian $h$ should not depend on 
$p_{0}.$ Consider, for instance, a fluid in adiabatic expansion. Its
instantaneous state is characterized, for a fixed particle number, by the
two variables $q^{1}\equiv U$ and $q^{2}\equiv V,$ and its equilibrium
properties by the entropy $q^{0}=S\left( U,V\right) .$ The intensive
variables are $\gamma _{1}=-p_{1}/p_{0}=1/T,\,\gamma _{2}=-p_{2}/p_{0}=P/T.$
If the motion of a piston changes the volume sufficiently slowly, according
to $\dot{V}\equiv \Phi ,$ the flux of energy is $\dot{U}=-P\Phi $ and the
resulting quasi-static dynamics of the fluid in the $6$-dimensional space $%
\mathbb{T}$ can be generated by the effective Hamiltonian 
\begin{equation}
h=\Phi \left[ p_{1}\frac{\partial S}{\partial V}\left/ \frac{\partial S}{%
\partial U}-p_{2}\right. \right]  \tag{4.16}
\end{equation}
(which possibly depends on time through $\Phi ).$ The constancy of entropy
is obvious from $\partial h/\partial p_{0}=0.$ In the $4$-dimensional
symplectic space $\overset{-}{\mathbb{T}}=U,V,\gamma _{1},\gamma _{2},$ the
resulting effective Hamiltonian is 
\begin{equation}
f=\Phi \left[ \gamma _{2}-\gamma _{1}\frac{\partial S}{\partial V}\left/ 
\frac{\partial S}{\partial U}\right. \right] ,  \tag{4.17}
\end{equation}
a simplified form of (4.14).

For \textit{continuous media,} the index $i$ in $q^{i}$ and $\gamma _{i}$
for $i\neq 0$ not only refers to the nature of the variable but also labels
the volume elements. This index thus includes the coordinates in ordinary
space, so that the variables $q^{i}$ and $\gamma _{i}$ constitute fields
(such as the energy density or the local velocity). The expression (4.14)
becomes the effective Hamiltonian for a canonical field theory. Moreover, in
hydrodynamics, the fluxes need not vanish with the gradients of the
intensive variables.

We have noted that, under the condition (4.2), the dynamics generated by the
Hamiltonian $h$ (or $f$) preserve the constraints (2.4) which express the
equations of state, provided they are satisfied at the initial time. One
could take advantage of these constraints so as to reduce the number of
variables, or to obtain by standard techniques non-canonical though
Hamiltonian dynamics in terms of reduced Poisson structures \cite{marsden}.

\section{Generation of equations of state}

We have written the equations of state which determine a thermodynamic
manifold $\mathbb{M}$ in the specific form (2.4). In practice a
thermodynamical system is characterized in the $2n+2$-dimensional space $%
\mathbb{T}$ by generalized equations of state, which constitute a set of $n+1
$ equations equivalent to the set (2.4). As a consequence of gauge
invariance, each of them is homogeneous in the $p_{i}$'s. These equations of
state are not always well known, and the missing ones may be determined
phenomenologically by comparison with known similar systems. To this aim,
contact transformations in the space $\overset{\sim }{\mathbb{T}}$ have been
used [8--11]. We suggest here to rely on canonical mappings or Hamiltonian
flows in the space $\mathbb{T}$ in order to transform the sets of equations
of state of different system into one another.

As an illustration, let us consider two systems characterized by their
entropy functions $S\left( q^{1},...,q^{n}\right) $ and $S^{\prime }\left(
q^{1},...,q^{n}\right) ,$ respectively. We wish to map their associated
thermodynamic manifolds $\mathbb{M}$ and $\mathbb{M}^{\prime }$ onto each
other in $\mathbb{T}$. For simplicity, we consider a mapping which does not
affect the coordinates $q^{1},...,q^{n}.$ The generating function $\mathcal{H%
}$ which achieves the canonical mapping through Eqs.\textrm{\ }(3.10) is not
defined in a unique fashion, since its action on manifolds other than $%
\mathbb{M}$ and $\mathbb{M}^{\prime }$ is not specified. We can readily
check that the following choice is suitable: 
\begin{equation}
\mathcal{H}=\sum\limits_{i=0}^{n}p_{i}^{\prime }q^{i}+p_{0}^{\prime }\left[
S^{\prime }\left( q^{1},...,q^{n}\right) -S\left( q^{1},...,q^{n}\right) %
\right] .  \tag{5.1}
\end{equation}%
Indeed, Eqs$\mathrm{.}$(3.10) yield 
\begin{eqnarray}
p_{0} &=&p_{0}^{\prime },\qquad p_{i}+p_{0}\frac{\partial S}{\partial q^{i}}%
=p_{i}^{\prime }+p_{0}^{\prime }\frac{\partial S^{\prime }}{\partial q^{i}},
\notag \\
q_{0}^{\prime }-S^{\prime } &=&q_{0}-S,\qquad q^{\prime i}=q^{i}\qquad
\left( i=1,...,n\right) ,  \TCItag{5.2}
\end{eqnarray}%
which are obviously satisfied for a pair of points located on $\mathbb{M}$
and $\mathbb{M}^{\prime }$ and having the same coordinates $%
p_{0},q^{1},...,q^{n}.$ Eq$\mathrm{.}$(5.1) exhibits the occurrence of a
thermodynamic potential, here the entropy, in the expression of the
generating function $\mathcal{H}$. Such an occurrence was already recognized
for contact transformations in $\overset{\sim }{\mathbb{T}}$ [9--11].

For a continuous set of hypothetical systems labelled by a deformation
parameter $\tau $ regarded as\ a mock time, the representative manifolds $%
\mathbb{M(}\tau )$ in $\mathbb{T}$ can be parametrized according to (2.4) by
means of the family $S_{\tau }\left( q^{1},...,q^{n}\right) $ of entropy
functions. The flow which transforms them into one another without changing
the coordinates $p_{0},q^{1},...,q^{n}$ is generated, according to (3.12)
and (5.1), by the time-dependent Hamiltonian 
\begin{equation}
h=p_{0}\frac{\partial }{\partial \tau }S_{\tau }\left(
q^{1},...,q^{n}\right) .  \tag{5.3}
\end{equation}%
More general flows transforming continuously the manifolds $\mathbb{M}\left(
\tau \right) $ into one another can be generated by Hamiltonians, obtained
by adding to (5.3) some $\tau $-dependent function which vanishes on $%
\mathbb{M}\left( \tau \right) $ and is homogeneous with degree 1 in the $%
p_{i}$'s. Indeed, as shown in Section 4, this additional term lets the
coordinates $p_{0},q^{1},...,q^{n}$ change in time\ without modifying $%
\mathbb{M}\left( \tau \right) .$

In the above example, we have assumed the entropy to be given as a
thermodynamic potential. For application purposes we can take advantage of
the geometric nature of the formalism, which allows us to parametrize the
thermodynamic manifolds with variables other than $q^{1},...,q^{n},p_{0}$
and accordingly to use different thermodynamic potentials. As an example,
let us reconsider transformations which map a van der Waals fluid onto
another one or onto a perfect gas, a problem already studied in the
framework of contact transformations \cite{benayoun1,benayoun2}. We keep
here the particle number fixed: we shall take for it the Avogadro number $N_{%
\mathrm{A}}.$ The space $\mathbb{T}$ has thus 6 dimensions. The practical
variables, namely, the molar entropy, energy and volume, the temperature,
the pressure and the single-particle chemical potential $\mu $ obtained from
(2.8) are identified as 
\begin{eqnarray}
S &=&q^{0},\qquad U=q^{1},\qquad V=q^{2},\qquad T=-\frac{p_{0}}{p_{1}}%
,\qquad P=\frac{p_{2}}{p_{1}},  \notag \\
\mu N_{\mathrm{A}} &=&U-TS+PV=\frac{1}{p_{1}}\left(
p_{0}q^{0}+p_{1}q^{1}+p_{2}q^{2}\right) .  \TCItag{5.4}
\end{eqnarray}%
The occurrence of the gauge factor $p_{1}$ in the denominators of the
intensive variables instead of $p_{0}$ is related to the use of the energy
representation, more convenient here than the entropy representation. In
fact, if we take the free energy $F\left( T,V\right) $ as a thermodynamic
potential instead of $S\left( U,V\right) ,$ we find the generalized
equations of state for a manifold $\mathbb{M}$ in the form 
\begin{equation}
q^{0}\equiv S=-\frac{\partial F}{\partial T},\qquad q^{1}+\frac{p_{0}}{p_{1}}%
q^{0}\equiv U-TS=F,\qquad \frac{p_{2}}{p_{1}}\equiv P=-\frac{\partial F}{%
\partial V},  \tag{5.5}
\end{equation}%
where the arguments in $F$ and its derivatives are replaced by $%
-p_{0}/p_{1}\equiv T$ and $q^{2}\equiv V.$ Here, $\mathbb{M}$ is thus
parametrized in terms of $q^{2},p_{0}$ and $p_{1}.$

A van der Waals fluid is characterized by its molar free energy 
\begin{equation}
F\left( T,V\right) =-RT\ln \left( V-b\right) -\frac{a}{V}+\Psi \left(
T\right) ,  \tag{5.6}
\end{equation}
where $R=N_{\mathrm{A}}k_{\mathrm{B}}$ and where the constants $a,b$ and the
function $\Psi \left( T\right) $ depend on the fluid.\ An ideal\ gas of
structureless particles with mass $m$ corresponds to the special case 
\begin{equation}
a=b=0,\qquad \Psi \left( T\right) =RT\ln \left[ \frac{N_{\mathrm{A}}}{%
\mathrm{e}}\left( \frac{2\pi \hslash ^{2}}{mk_{\mathrm{B}}T}\right) ^{3/2}%
\right] .  \tag{5.7}
\end{equation}

We wish to map this van der Waals fluid onto another one (or onto an ideal
gas), characterized by its free energy $F^{\prime }\left( T,V\right) $ of
the form (5.6) where $a,b,\Psi \left( T\right) $ are replaced by $a^{\prime
},b^{\prime },\Psi ^{\prime }\left( T\right) .$ We consider in $\mathbb{T}$
a mapping of $\mathbb{M}$ onto $\mathbb{M}^{\prime }$ for which the
coordinates $q^{2},p_{0}$ and $p_{1},$ that is, $T$ and $V,$ remain
unchanged. This is achieved by introducing a generating function $\mathcal{K}%
,$ the Legendre transform of $\mathcal{H}$ with respect to $p_{1}^{\prime }$
and $q^{1},$ due to the choice of variables $T$ and $V$ in the thermodynamic
potential. Transposing Eq.(5.1) yields 
\begin{equation}
\mathcal{K}\!\!\left( q^{0},p_{1},q^{2},p_{0}^{\prime },q^{\prime
1},p_{2}^{\prime }\right) \!\!=p_{0}^{\prime }q^{0}\!-\!p_{1}q^{\prime
1}\!+\!p_{2}^{\prime }q^{2}\!+\!p_{1}\!\!\left[ \!F^{\prime }\!\left( -\frac{%
p_{0}^{\prime }}{p_{1}},q^{2}\right) \!\!-\!\!F\!\!\left( -\frac{%
p_{0}^{\prime }}{p_{1}},q^{2}\right) \!\!\right] \!\!,  \tag{5.8}
\end{equation}
where $F$ and $F^{\prime }$ have the form (5.6) or (5.7). We readily check
that the mapping provided by the partial derivatives of $\mathcal{K}$ in the
full space $\mathbb{T}$ reads: 
\begin{eqnarray}
\!\!\!\!p_{0} &=&p_{0}^{\prime },\quad p_{1}^{\prime }=p_{1},\quad
p_{2}+p_{1}\frac{\partial F}{\partial V}=p_{2}^{\prime }+p_{1}^{\prime }%
\frac{\partial F^{\prime }}{\partial V},\quad q^{\prime 2}=q^{2},  \notag \\
q^{0}+\frac{\partial F}{\partial T} &=&q^{\prime 0}+\frac{\partial F^{\prime
}}{\partial T},\quad q^{1}-\frac{p_{0}}{p_{1}}\frac{\partial F}{\partial T}%
-F=q^{\prime 1}-\frac{p_{0}^{\prime }}{p_{1}^{\prime }}\frac{\partial
F^{\prime }}{\partial T}-F^{\prime }.  \TCItag{5.9}
\end{eqnarray}
Hence the equations of state (5.5) for $\mathbb{M}$ imply the corresponding
ones for $\mathbb{M}^{\prime },$ and $\mathcal{K}$ therefore maps as
expected these manifolds onto each other.

If $a,b,\Psi \left( T\right) $ depend continuously on a mock time $\tau ,$
the free energy (5.6) depends on $\tau ,$ what we denote by $F_{\tau }.$ The
generating function $\mathcal{K}$ of (5.8) taken between the times $\tau $
and $d\tau $ yields a time-dependent Hamiltonian, as $\mathcal{H}$ does in
(3.12). The family of manifolds $\mathbb{M}\left( \tau \right) $ is then
obtained from the Hamiltonian flow produced by 
\begin{equation}
h=p_{1}\frac{\partial }{\partial \tau }F\left( -\frac{p_{0}}{p_{1}}%
,q^{2}\right) .  \tag{5.10}
\end{equation}
The explicit form of $h$ follows from (5.6).

In the above mappings through (5.8) or (5.10) the variables $%
q^{2},p_{0},p_{1}$ and accordingly $V,T$ are taken as\ constants of the
motion. We can also build Hamiltonian mappings which modify $V$ or $T.$ As a
simple example, let us construct a transformation among the set of van der
Waals equations which will make the law of corresponding states obvious. The
critical point is given by 
\begin{equation}
V_{\mathrm{c}}=3b,\qquad P_{\mathrm{c}}=\frac{a}{27b^{2}},\qquad T_{\mathrm{c%
}}=\frac{8a}{27b},  \tag{5.11}
\end{equation}%
and the $P,V,T$ equation of state depends only on the reduced variables $%
V/V_{\mathrm{c}},P/$ $P_{\mathrm{c}},T/T_{\mathrm{c}}.$ We look for a
mapping which leaves these reduced variables invariant. If moreover we wish $%
T\equiv -p_{0}/p_{1}$ and hence $T_{\mathrm{c}}$ to remain constant in time,
we should find that the variables $V\equiv q^{2}$ and $P^{-1}\equiv
p_{1}/p_{2}$ as well as the parameters $a$ and $b$ in the equation of state
vary proportionally to one another. On the other hand, the variations of $V$
and $b$ entail that the contribution $R\ln \left( V-b\right) $ to the
entropy $S\equiv q^{0}$ depends on the time $\tau .$ If we assume this
dependence to be linear, in $-R\tau ,$ the common dependence of $V=q^{2},b,$
and $P^{-1}=p_{1}/p_{2}$ should be exponential, in $\mathrm{e}^{-\tau }.$
The simplest Hamiltonian which achieves these goals is 
\begin{equation}
h=-Rp_{0}-p_{2}q^{2},  \tag{5.12}
\end{equation}%
since it provides the equations of motion 
\begin{equation}
\frac{dp_{0}}{d\tau }=\frac{dp_{1}}{d\tau }=0,\;\frac{dp_{2}}{d\tau }%
=p_{2},\;\frac{dq^{0}}{d\tau }=-R,\;\frac{dq^{1}}{d\tau }=0,\;\frac{dq^{2}}{%
d\tau }=-q^{2},  \tag{5.13}
\end{equation}%
implying $V\propto P^{-1}\propto \mathrm{e}^{-\tau }$ and constant $T.$ The
left-hand side of the $P,V,T$ equation of state, which is at the initial
time 
\begin{equation}
\left( P+\frac{a}{V^{2}}\right) \left( V-b\right) -RT=0,  \tag{5.14}
\end{equation}%
retains as expected its form in this motion of $P$ and $V,$ with $a$ and $b$
changed into $a\mathrm{e}^{-\tau }$ and $b\mathrm{e}^{-\tau }.$ The
remaining two equations of state (5.5), which involve $S\equiv q^{0}$ and $%
U\equiv q^{1},$ also retain their form, and the choice (5.12) for the
Hamiltonian entails that the function $\Psi \left( T\right) $ does not
depend on $\tau .$ Adding to the Hamiltonian (5.12) a homogeneous function
of $p_{0}$ and $p_{1}$ with degree 1 would result in a change of $\Psi
\left( T\right) $ in the free energy (5.6).

The molar chemical potential $\mu N_{\mathrm{A}}$ obtained from (5.4) and
(5.13) varies linearly as $RT\tau $ along the trajectories.\ If we consider
the two end points of the plateau in the isotherms which describes
liquid-vapour equilibrium, the equality of the intensive variables $T,P,\mu $
at these points is maintained along the trajectories, so that the flow maps
the saturation curve into that of the transformed fluid. This is consistent
with Maxwell's construction and with the conservation of areas in the $P,V$
plane in a symplectic flow with constant $T.$ Note finally that the ideal
gas equation of state remains unchanged in the mock dynamics generated by
(5.12), and that a van der Waals fluid is changed into an ideal gas by this
dynamics in the large $\tau $ limit.

\section{Metric structure}

We have not yet dealt with the concavity of the entropy $S\left(
q^{1},...,q^{n}\right) $ as function of the extensive variables, which
expresses the stability of equilibrium states. This property produces
constraints on the physical manifolds $\mathbb{M}$ in the $2n+2$-dimensional
space, that we wish to express. It entails the positivity of the matrix $%
-\partial ^{2}S/\partial q^{i}\partial q^{j},$ or equivalently the existence
of a metric structure in the $n$-dimensional space $q^{i}$ relying on the
quadratic form 
\begin{equation}
ds^{2}=-d^{2}S=-\sum\limits_{i,j=1}^{n}\frac{\partial ^{2}S}{\partial
q^{i}\partial q^{j}}dq^{i}\,dq^{j},  \tag{6.1}
\end{equation}%
which defines a distance between two neighbouring thermodynamic states.

The possibility of endowing equilibrium thermodynamics with some Riemannian
structure is known since a long time \cite{weinhold1, ruppeiner}, but the
metric (6.1) is not the only possible one in the framework of macroscopic
thermodynamics. For instance, $ds^{2}=d^{2}U$ also appears acceptable,
although $ds^{2}=-d^{2}S$ is preferable in the framework of thermodynamic
fluctuation theory \cite{ruppeiner}.

This arbitrariness can be lifted by resorting to statistical mechanics. The
problem there is to find a physically meaningful distance \cite{balian2}
between two arbitrary neigbouring density operators $\hat{D}$ and $\hat{D}+d%
\hat{D}.$ To this aim one should rely on the formal structure of quantum
statistical mechanics. The essential fact is that the only quantities which
behave as scalars in a change of representation and which have physical
relevance at a givent time are (i) the expectation values $\left\langle
A\right\rangle =\mathrm{Tr}\hat{D}\hat{A}$ of the Hermitean operators $%
\hat{A}$ in Hilbert space which represent any physical variable, and (ii)
the von Neumann entropy 
\begin{equation}
S_{\mathrm{vN}}=-k_{\mathrm{B}}\mathrm{Tr}\hat{D}\ln \hat{D}.  \tag{6.2}
\end{equation}%
Hence the operators $\hat{D}$ on the one hand and $\hat{A}$\ or $\ln \hat{D}$%
\ on the other hand can be regarded as belonging to two dual vector spaces.
From the above considerations one can show \cite{balian2} that the only
quadratic form in $d\hat{D}$\ which has physical meaning is, within a
multiplicative constant,%
\begin{equation}
ds^{2}=-d^{2}S_{\mathrm{vN}}=k_{\mathrm{B}}\mathrm{Tr}\left[ \left( d\hat{D}%
\right) \left( d\ln \hat{D}\right) \right] .  \tag{6.3}
\end{equation}%
It is positive owing to the concavity of the von Neumann entropy and
therefore defines a natural Riemannian metric in the space of density
operators $\hat{D}.$ The matrix elements of $d\hat{D}$\ and $k_{\mathrm{B}%
}d\ln \hat{D}$\ appear as the contravariant and the covariant coordinates of
a shift. Returning to thermodynamics, we can restrict this metric to the
subset of the space $\hat{D}$\ constituted by the Gibbsian density
operators, which have a form of the type (2.1) and which represent at the
microscopic scale the local equilibrium states of thermodynamics.\ These
density operators are in one-to-one correspondence with the points $%
q^{1},...,q^{n}$ and their von Neumann entropy (6.2) is identified with the
entropy $S\left( q^{1},...,q^{n}\right) $ of thermodynamics. The reduction
of the metric (6.3) to this $n$-dimensional subset\ of the space $\hat{D}$
is identical with the metric (6.1), which\ thus arises as the \textit{only
natural metric }inferred on equilibrium or non-equilibrium\textit{\ }%
thermodynamics by the unique metric of\ quantum statistical physics.

In this microscopic approach the volume of the system is regarded as fixed.
It is not included in the variables $q^{i}$ which characterize the state of
the system. If the set $q^{i}$ contains all the extensive variables,
including the volume, an eigenvalue $0$ appears in the metric tensor (6.1),
since the extensivity of $S$ then implies 
\begin{equation}
\sum\limits_{j=1}^{n}\frac{\partial ^{2}S}{\partial q^{i}\,\partial q^{j}}%
q^{j}=0.  \tag{6.4}
\end{equation}%
The distance (6.1) between two states also vanishes in the thermodynamic
limit in a phase equilibrium situation, when these two states differ only
through the proportion of the phases. Distances are always positive between
two states having different intensive variables, and they thus characterize
their qualitative differences.

The condition $ds^{2}\geq 0$ expresses one of the Laws of thermodynamics,
the concavity of the entropy function. In order to complete the formulation
or thermodynamics in the spaces $\overset{-}{\mathbb{T}},\overset{\sim }{%
\mathbb{T}}$ or $\mathbb{T},$ we therefore have to rewrite (6.1) in terms of
points on a manifold $\overset{-}{\mathbb{M}},\overset{\sim }{\mathbb{M}}$
or $\mathbb{M}$. We first note that, for a given manifold $\overset{-}{%
\mathbb{M}},$ the definition (1.1) of the intensive variables $\gamma _{i}$
implies 
\begin{equation}
d\gamma _{i}=\sum\limits_{j=1}^{n}\frac{\partial ^{2}S}{\partial
q^{i}\,\partial q^{j}}dq^{j}.  \tag{6.5}
\end{equation}%
Thus, if we parametrize the $n$-dimensional manifold $\overset{-}{\mathbb{M}}
$ associated with a given system in the $2n$-dimensional configuration space 
$\overset{-}{\mathbb{T}}$ by means of the coordinates $q^{1},...,q^{n},$ the
metric (6.1) takes a simple form: An infinitesimal shift of the state can be
represented either by the set of variations $dq^{1},...,dq^{n}$ which can be
regarded as its \textit{contravariant} components, or by the set $-d\gamma
_{1},...,-d\gamma _{n}$\ which according to Eq. (6.5), appear as its \textit{%
covariant} ones. The distance between two states of any given physical
system takes, for infinitesimal variations on $\mathbb{\bar{M}},$\ the form 
\begin{equation}
ds^{2}=-\sum\limits_{i=1}^{n}d\gamma _{i}\,dq^{i},  \tag{6.6}
\end{equation}%
which does not depend explicitly on the entropy function of the system. The
specificity of this system is reflected by the curvature of the manifold $%
\overset{-}{\mathbb{M}},$ itself resulting from the metric tensor given by
(6.5).

Considered in the whole space $\overset{-}{\mathbb{T}},$ the expression
(6.6), which equivalently reads 
\begin{equation*}
ds^{2}=\sum\limits_{i=1}^{n}\left[ \frac{1}{2}\left( dq^{i}-d\gamma
_{i}\right) \right] ^{2}-\left[ \frac{1}{2}\left( dq^{i}+d\gamma _{i}\right) %
\right] ^{2},
\end{equation*}%
defines a pseudo-Euclidean metric with signature $\left( n,n\right) .$ It
may have either sign between two neighbouring points of $\overset{-}{\mathbb{%
T}}$ which correspond to different physical systems.\ However, the
restriction of (6.6) to a thermodynamic manifold $\overset{-}{\mathbb{M}}$
defines the positive Riemannian metric (6.1). Conversely, the submanifolds $%
\overset{-}{\mathbb{M}}$ of $\overset{-}{\mathbb{T}}$ which can be
thermodynamically admissible are the \textit{Lagrangean manifolds} over
which (6.6) moreover \textit{induces a positive Riemannian metric.} The
first condition ensures the existence of a function $S$ which generates the
equations of $\overset{-}{\mathbb{M}}$ in the form (1.1), and the second one
ensures the concavity of this function.

The pseudo-Euclidean metric (6.6) also holds for the $2n+1$-dimensional
thermodynamic space $\overset{\sim }{\mathbb{T}}.$ The positivity of the
metric that it induces on a $n$-dimensional \textit{Legendrean} submanifold $%
\overset{\sim }{\mathbb{M}}$ of $\overset{\sim }{\mathbb{T}}$ now ensures
that this manifold may represent a thermodynamically stable physical system.

Similar features occur in the $2n+2$-dimensional space $\mathbb{T}.$ The
pseudo-Euclidean metric (6.6) can be rewritten as 
\begin{equation}
ds^{2}=\frac{1}{p_{0}}\sum\limits_{i=0}^{n}dp_{i}\,dq^{i}.  \tag{6.7}
\end{equation}%
The factor $1/p_{0}$ ensures gauge invariance, and apart from it $ds^{2}$ is
symmetric with respect to $i=0,1,...,n.$ The restriction of (6.7) to a $n+1$%
-dimensional Lagrangean manifold $\mathbb{M}$ where $\omega \equiv
\sum\limits_{i=0}^{n}p_{i}\,dq^{i}=0$ (which ensures the existence of $S$
and the gauge invariance) yields again the metric (6.1) on $\mathbb{M}.$
However, there are here $n+1$ contravariant coordinates $%
dq^{0},dq^{1},...,dq^{n}$ for a shift in $\mathbb{M},$ which are not
independent since $\omega =0.$ The independent variables on $\mathbb{M}$
being taken, for instance,$\;$as $q^{1},...,q^{n},p_{0},$ the covariant
coordinates of a shift are $dp_{i}/p_{0}\;(i=0,1,...,n).$ The distance
between two points describing the same state, which are deduced from each
other through a gauge transformation (2.3), vanishes. More generally two
different states with the same intensive variables $\gamma _{1},...,\gamma
_{n}$ lie at a vanishing distance, as obvious from (6.6) for $\overset{-}{%
\mathbb{M}}$ or $\overset{\sim }{\mathbb{M}}$ and from (6.7) in a gauge with
constant $p_{0}$ for $\mathbb{M};$ this expresses that the metric
characterizes the nature of the independent phases of the system, not their
sizes.

Altogether the physical manifolds $\mathbb{M}$ in $\mathbb{T}$ should not
only satisfy the condition $\omega =0,$ which ensures that the thermodynamic
identities are satisfied, but should also be such that the bilinear
differential form (6.7) \textit{induces on them a non}-\textit{negative
Riemannian metric} $ds^{2},$ which ensures thermodynamic stability.

Accordingly, canonical transformations $\mathcal{H}$ which map $\mathbb{M}$
onto $\mathbb{M}^{\prime }$ in $\mathbb{T}$ preserve thermodynamic stability
only if they satisfy some inequalities. Using Eqs. (3.8), we can compare the
metrics at corresponding points through 
\begin{eqnarray}
ds^{2} &=&\left( \frac{\partial \mathcal{H}}{\partial q^{0}}\right)
^{-1}\sum\limits_{i,j=0}^{n}\left( \frac{\partial ^{2}\mathcal{H}}{\partial
p_{i}^{\prime }\,\partial q^{j}}dp_{i}^{\prime }\,dq^{j}+\frac{\partial ^{2}%
\mathcal{H}}{\partial q^{i}\,\partial q^{j}}dq^{i}\,dq^{j}\right) ,  \notag
\\
ds^{\prime 2} &=&\frac{1}{p_{0}^{\prime }}\sum\limits_{i,j=0}^{n}\left( 
\frac{\partial ^{2}\mathcal{H}}{\partial p_{i}^{\prime }\,\partial q^{j}}%
dp_{i}^{\prime }\,dq^{j}+\frac{\partial ^{2}\mathcal{H}}{\partial
p_{i}^{\prime }\,\partial p_{j}^{\prime }}dp_{i}^{\prime }\,dp^{\prime
j}\right) .  \TCItag{6.8}
\end{eqnarray}%
The mapping should be such that, if the points $q^{i},p_{i}$ and $q^{\prime
i},p_{i}^{\prime }\left( i=0,1,...,n\right) $ and their variations lie on
the manifolds $\mathbb{M}$ and $\mathbb{M}^{\prime },$ respectively, both $%
ds^{2}$ and $ds^{\prime 2}$ are positive. In a continuous transformation
generated by $h,$ the metric evolves according to 
\begin{equation}
\frac{d}{d\tau }\left( ds^{2}\right) =\frac{1}{p_{0}}\frac{\partial h}{%
\partial q^{0}}ds^{2}+\frac{1}{p_{0}}\sum\limits_{i,j=0}^{n}\left( \frac{%
\partial ^{2}h}{\partial q^{i}\,\partial p_{j}}dp_{i}\,dp_{j}-\frac{\partial
^{2}h}{\partial q^{i}\partial q^{j}}dq^{i}dq^{j}\right) .  \tag{6.9}
\end{equation}%
A control of the sign of the metric through Eqs.\ (6.8) or\ (6.9) sets up
conditions on $\mathcal{H}$ or $h.$ The occurrence of a zero eigenvalue in
the metric tensor during the evolution in terms of the mock time $\tau $
indicates the appearance of a critical point in the equations of state thus
generated.

\section{Conclusion}

The extension of the $n$-dimensional configuration space $q^{1},...,q^{n}$
associated with a thermodynamic system, successively to the $2n$-dimensional
space $\overset{-}{\mathbb{T}}$ including the intensive variables $\gamma
_{1},...,\gamma _{n},$ then to the $2n+1$-dimensional space $\overset{-}{%
\mathbb{T}}$ including in addition the entropy $q^{0},$ finally to the $2n+2$%
-dimensional space $\mathbb{T}$ where the $\gamma _{i}$'s are replaced by
the variables $p_{0},p_{1},...,p_{n}$ including a gauge arbitrariness,
presents practical advantages. It is adapted\ to changes of variables\ among
the various physical quantities. It sets into the same framework all the
systems which have the same degrees of freedom $q^{1},...,q^{n}$ but
different equations of state. Moreover it discloses a rich mathematical
structure that characterizes the geometry of the spaces $\overset{-}{\mathbb{%
T}},$ $\overset{\sim }{\mathbb{T}}$ or $\mathbb{T},$ and of the manifolds $%
\overset{-}{\mathbb{M}},\overset{\sim }{\mathbb{M}}$ or $\mathbb{M}$ which
represent a thermodynamic system.

The existence of thermodynamic identities is reflected in the Legendre
structure of any $\overset{\sim }{\mathbb{M}}$, characterized by the
identity $\tilde{\omega}\equiv dq^{0}-\sum\limits_{i=1}^{n}\gamma
_{i}dq^{i}=0,$ or equivalently in the Lagrange structure of any $\overset{-}{%
\mathbb{M}}$ or $\mathbb{M}$ characterized by the more symmetric identity $%
\omega \equiv \sum\limits_{i=0}^{n}p_{i}dq^{i}=0.$ The conservation of $%
\omega $ in transformations which map the manifolds $\mathbb{M}$ onto one
another induces for $\mathbb{T}$ a symplectic structure based on the 2-form $%
d\omega =\sum\limits_{i=0}^{n}dp_{i}\wedge dq^{i},$ where the variables $%
q^{i}$ and $p_{i}$ appear as conjugate in the sense of analytical mechanics.
Such mappings are then canonical, and can be generated by Hamiltonian flows.
The quantities $q^{i}$ play the r\^{o}le of position variables, the
quantities $p_{i}$ or $\gamma _{i}$ that of momentum variables. The
Hamiltonians involved should be homogeneous with degree 1 in the variables $%
p_{i}$ so as to preserve the gauge invariance of the manifolds $\mathbb{M}.$
The mappings such that the trajectory issued from any point of some given
manifold $\mathbb{M}_{0}$ lies in $\mathbb{M}_{0}$ are those for which the
Hamiltonian vanishes over $\mathbb{M}_{0}.$\ Finally the stability of
thermodynamic states is reflected in the fact that the bilinear forms $%
-\sum\limits_{i=1}^{n}d\gamma _{i}\,dq^{i}$ or $p_{0}^{-1}\sum%
\limits_{i=0}^{n}dp_{i}\,dq^{i}$ should induce on $\overset{-}{\mathbb{M}},%
\overset{\sim }{\mathbb{M}}$ or $\mathbb{M}$ a Riemannian structure with a
non-negative metric.

We have shown how Hamiltonians can be constructed, either for real motions
in thermodynamic processes, even when they are dissipative, or in fictitious
dynamics which consistently generate equations of state for different
systems.

The above considerations should have more than a formal value.\ We have
worked out a few elementary examples for the sole purpose of illustration.
More realistic problems should benefit from the existence of an ordinary
Hamiltonian structure, which for instance readily generates variational
techniques, and which anyhow is formally simpler than the currently used
contact structure.

\vskip 0.3cm%

{\small We wish to thank the Total\thinspace Fina\thinspace Elf\thinspace
group for support and permission to publish. We are also grateful to the
referee for his thorough screening of the manuscript.r }

\vskip 0.3cm%


\begin{thebibliography}{99}
\bibitem{callen1} H.B. Callen,\ \textit{Thermodynamics and an introduction
to Thermostatistics }(Wiley, 1985).

\bibitem{balian1} see, for instance, R. Balian, \textit{From Microphysics to
Macrophysics, Methods and Applications of Statistical Physics, }Vols. I and
II (Springer Verlag, 1991, 1992).

\bibitem{balian2} R. Balian, Y.\ Alhassid and H.\ Reinhardt, Phys.\ Reports\ 
\textbf{131}, 1 (1986).

\bibitem{rau} J.\ Rau and B.\ M\"{u}ller, Phys.\ Reports\ \textbf{272}, 1
(1996).

\bibitem{caratheodory} C. Caratheodory, \textit{Calculus of variations and
partial differential equations of the first order}, Vol.\ I (Holden-Day, San
Francisco, 1965).

\bibitem{hermann} R. Hermann, \textit{Geometry, Physics and Systems }%
(Dekker, New York, 1973).

\bibitem{arnold1} V.I. Arnold, in \textit{Proceedings of the 150th
anniversary Gibbs symposium} (New Haven, 1989), p. 163 (Amer. Math. Soc.,
1990).

\bibitem{mrugala} R. Mrugala, J. D. Nulton, J. C. Schoen, and P. Salamon,\
Phys. Rev. \textbf{A} \textbf{41}, 3156 (1990) and Rep.\ Math.\ Phys\textit{%
., }\textbf{29}, 109 (1991).

\bibitem{benayoun1} L. Benayoun,\ \textit{M\'{e}thodes g\'{e}om\'{e}triques
pour l'\'{e}tude des syst\`{e}mes thermodynamiques et la g\'{e}n\'{e}ration
d'\'{e}quations d'\'{e}tat. }Thesis, Institut National Polytechnique de
Grenoble, 1999.

\bibitem{mrugala2} R. Mrugala, Rep.\ Math.\ Phys.\ \textbf{33}, 149 (1993);
and in \textit{Thermodynamics of energy conversion and transport,} edited by
S. Sieniutycz and A. De Vos, p. 276 (Springer, 2000).

\bibitem{benayoun2} L. Benayoun and P.\ Valentin (To be published).

\bibitem{herglotz1} G. Herglotz, \textit{Lectures on Contact Transformations
and Hamiltonian systems} (1932), edited by R.B. Guenther, C. M. Guenther, J.
A. Gottsch, \textit{Lecture Notes in Nonlinear analysis (}N. Copernicus
University, Torun, 1996).

\bibitem{arnold2} V. I. Arnold, \textit{Mathematical Methods of Classical
Mechanics} (Springer, 1989).

\bibitem{kijowski1} J. Kijowski and W. Tulczyjew, \textit{A symplectic
framework for field theories,} Lecture notes in physics 107 (Springer, 1979).

\bibitem{omohundro1} S. M. Omohundro, \textit{Geometric Perturbation Theory
in Physics} (World Scientific 1986), Chap.\ 16.

\bibitem{janeczko} S. Janeczko, Ann. Soc.\ Sci. Bruxelles, \textbf{99}, 49
(1985).

\bibitem{morrison1} P. J. Morrison, Rev.\ Mod.\ Phys.\ \textbf{70}, 467
(1998).

\bibitem{weinhold1} F. Weinhold, J.\ Chem.\ Phys.\ \textbf{63}, 2479, 2484,
2488, 2496 (1973) and \textbf{65}, 559 (1975).

\bibitem{ruppeiner} G. Ruppeiner, Rev.\ Mod.\ Phys.\ \textbf{67}, 605 (1995)
and enclosed references.

\bibitem{landau} L. D. Landau and E. M. Lifshitz, \textit{Mechanics}
(Pergamon, 1976).

\bibitem{marsden} see the contributions of J. E. Marsden, P. J. Morrison and
A. Weinstein, in \textit{Fluids and plasmas : Geometry and dynamics}, edited
by J. E. Marsden, Contemporary Maths., Vol. 28 (Am.\ Math.\ Soc.,\
Providence, 1984).
\end{thebibliography}
\end{document}